\documentclass[aps,reprint,superscriptaddress]{revtex4-2}
\usepackage{graphicx}
\usepackage{dcolumn}
\usepackage{bm}
\usepackage{color}
\usepackage{xcolor}
\usepackage{natbib}
\usepackage[utf8]{inputenc}
\usepackage[T1]{fontenc}
\usepackage{hyperref}
\usepackage{float}
\usepackage{amsfonts}
\usepackage{textgreek} 

\usepackage{multirow}

\makeatletter
\def\maketitle{
\@author@finish
\title@column\titleblock@produce
\suppressfloats[t]}
\makeatother

\newcommand{\var}{\textrm{var}}
\newcommand{\std}{\textrm{std}}

\begin{document}



\author{Yann Lanoisel\'ee}
\email{Corresponding author: ylanoiselee@bcamath.org }
\affiliation{
BCAM -- Basque Center for Applied Mathematics,
Alameda de Mazarredo 14, 48009 Bilbao, Basque Country -- Spain}

\author{Gianni Pagnini}
\email{Contact author: gpagnini@bcamath.org }
\affiliation{
BCAM -- Basque Center for Applied Mathematics,
Alameda de Mazarredo 14, 48009 Bilbao, Basque Country -- Spain}
\affiliation{
Ikerbasque -- Basque Foundation for Science, Plaza Euskadi 5, 48009 Bilbao, Basque Country -- Spain}

\author{Agnieszka Wyłomańska}
\email{Contact author: agnieszka.wylomanska@pwr.edu.pl}
\affiliation{ 
Faculty of Pure and Applied Mathematics, Hugo Steinhaus Center, Wroclaw University of Science and Technology, Wyspianskiego 27, 50-370 Wroclaw, Poland}

\begin{abstract}
The molecular motion in heterogeneous media displays anomalous diffusion by the mean-squared displacement 
$\langle X^2(t) \rangle = 2 D t^\alpha$. 
Motivated by experiments reporting populations of the anomalous diffusion parameters $\alpha$ and $D$, we aim to disentangle their respective contributions to the observed variability when this last is due to a true population of these parameters and when it arises due to finite-duration recordings.
We introduce estimators of the anomalous diffusion parameters on the basis of the time-averaged 
mean-squared displacement and study their statistical properties.
By using a copula approach, we derive a formula for the joint density function of their estimations conditioned on their actual values. The methodology introduced is indeed universal, it is valid for any Gaussian process and can be applied to any quadratic time-averaged statistics. 
We also explain the experimentally reported relation 
$D \propto \exp(\alpha c_1+c_2)$ for which we provide the exact expression.
We compare our findings to numerical simulations of the fractional Brownian motion
and quantify their accuracy by using the Hellinger distance. Finally, we provide guidelines and 
routines for fitting experimental data.
\end{abstract}
 
\date{\today}
\title{Super-resolved anomalous diffusion:\\
deciphering the joint distribution of anomalous exponent and diffusion coefficient}
\maketitle

Diffusion in heterogeneous media is relevant to physics, chemistry, and biology. Particle dynamics in a heterogeneous system reflect the interaction between a particle and its environment. 
Physical features such as reaction rates and ergodicity breaking, as well as statistical features, among which rare events and non-Markovianity, are intrinsically connected to the nature of medium heterogeneity. Recent single-particle tracking experiments in heterogeneous and crowded environments, for instance biological cells, have reported anomalous diffusion \cite{Banks2005,Szymanski2009,Bronstein2009,Hofling_2013}. 
In brief, a diffusive process $X(t)$ is labeled as anomalous if its ensemble-averaged mean-squared displacement (EAMSD) grows in time as a power-law function, i.e., 
\begin{equation}\label{eq:MSD}
 \langle X^2(t)\rangle=2Dt^{\alpha} \,,
\end{equation}
where $\alpha\in[0,2]$ is the anomalous exponent and $D\in[0,\infty)$ is the generalized diffusion coefficient (hereinafter called diffusion coefficient) with units $m^2s^{-\alpha}$
, and $\langle \cdot\rangle$ denotes the expectation value.
Anomalous exponent $\alpha$ can vary due to a change in the viscoelastic properties of the medium or to regime-switching between passive and active motion \cite{Arcizet2008}. Similarly, the diffusion coefficient $D$ can be affected by changes in viscosity, hydrodynamic radius, or even temperature. 


The distribution of the anomalous exponents has been discussed in the context of histone-like nucleoid-structuring proteins dynamics~\cite{wang2018}, and quantum dots in cell cytoplasm~\cite{sabri2020elucidating,janczura2021identifying,cherstvy2019non}. The distribution of the diffusion coefficients has been discussed in $\beta$-adrenergic receptor dynamics~\cite{Grimes2023}. The joint distribution of anomalous exponent and diffusion coefficient has been studied in the diffusion of G proteins \cite{Sungkaworn2017}, intracellular quantum dots \cite{Etoc2018},
$\kappa$-opioid receptor~\cite{Drakopoulos2020}, membrane-less organelles in 
\textit{C. elegans} embryos \cite{benelli2021sub}, nanobeads in clawfrog \textit{X. laevis} egg extract~\cite{speckner2021single}, and endosomal dynamics~\cite{korabel_etal-e-2021}. Additionally, the correlation between anomalous exponent and diffusion coefficient shows, in 
some cases, similar patterns~\cite{cherstvy2019non,speckner2021single,benelli2021sub,korabel_etal-e-2021} for which we provide here an explanation.


Since the analysis of individual trajectories from experimental data suggests that $\alpha$ and $D$ can be randomly distributed, 
we address here the problem of characterizing the conditional joint distribution of the estimated parameters $(\hat{\alpha} \,, \hat{D})$ 
given the expected pair $(\alpha \,, D)$. From the experimental point of view, the stochastic nature of the system and the finite duration of the experimental trajectories introduce errors in the estimation of the parameters. For an expected pair $(\alpha \,, D)$, each realization of the process yields values of the estimated parameters $(\hat{\alpha} \,, \hat{D})$ affected by random fluctuations and resulting in a probability density function (PDF) $p(\hat{\alpha},\hat{D})$. Shorter trajectories exacerbate these estimation errors. This introduces a `resolution limit' to the joint PDF of the estimated parameters. In fact, when the true physical parameters are randomly distributed, the joint PDF of the estimated parameters can be viewed as a convolution between the distribution of the physical parameters and that of the estimation errors. The additional difficulty here is that the error level itself depends on $\alpha$ and $D$. Moreover, $\hat{\alpha}$ and $\hat{D}$ are correlated due to the formulation of the mathematical estimators.

Theoretically, this is the framework of superstatistics \cite{beck_etal-pa-2003} 
 where, in addition to thermal noise, further randomness is provided by fluctuations of intensive quantities along particles' trajectories or by distinct parameters' values for each particle. Thus, we enable the identification of potential superstatistical anomalous diffusion models where both $\alpha$ and $D$ can be randomly distributed even in the case of short trajectories. Furthermore, since the fractional Brownian motion (fBm) turned out to be a good candidate for the underlying stochastic motion in many living systems \cite{magdziarz_etal-prl-2009,szymanski_etal-prl-2009,weiss-pre-2013}, the superstatistical fBm is considered in this study as a prototypical case. Superstatistical fBm characterized by a population of diffusion coefficients has been formulated within the framework of the generalized grey Brownian motion \cite{mura_etal-jpa-2008,pagnini_etal-fcaa-2016} and also checked against experimental literature \cite{mackala_etal-pre-2019,runfola_etal-rsos-2022}. More mathematical aspects related to ergodicity breaking \cite{molina_etal-pre-2016} and the generalized Fokker--Planck equation were also studied \cite{pagnini-fcaa-2012,runfola_etal-pd-2024}. Additionally, fBm with a random anomalous exponent (called Hurst exponent) was also studied \cite{Balcerek2022,han2020deciphering,korabel2021local,Balcerek_2023,PhysRevResearch.5.L032025}, and intriguing phenomena such as accelerated diffusion and time-dependent persistence transitions were observed. Extensions of fBm, where the Hurst exponent behaves as a stationary random process, have also been discussed in mathematical studies \cite{levy1995multifractional,ayachetaqqu2005,PhysRevLett.134.197101}. 
 Finally, a two-variable superstatistical fBm has been
preliminary discussed in the literature
\cite{itto_etal-jrsi-2021}. 

Thus, in the following, first, we propose a general methodology for identifying the joint PDF of true parameters $p(\alpha, D)$ given the joint PDF of the estimated ones $p(\hat{\alpha}, \hat{D})$ and, later, while our approach is valid for any ergodic Gaussian process, we show it at work in the case of the fBm \cite{beran2016long}. We also discuss the accuracy of the approach as a function of trajectory length. We conclude by reporting fitting guidelines and routines in the End Matter~\ref{sec:End_Matter}.

The joint PDF $p(\hat{\alpha},\hat{D})$
is determined by 
\begin{equation}\label{eq:joint_pdf_general_case}
p(\hat{\alpha},\hat{D})=\int_0^2d\alpha\int_0^\infty dD\, p(\hat{\alpha},\hat{D}|\alpha,D) p(\alpha,D)
\,,
\end{equation}
where $p(\hat{\alpha},\hat{D}|\alpha,D)$ is the joint conditional PDF of the estimated anomalous exponent and diffusion coefficient given the true values $\alpha$ and $D$, respectively, and we call it the 'transfer function'. Note that the transfer function depends implicitly on the trajectory length and the lag times used in the fitting.
When parameters
$\alpha$ and $D$ take a single value, 
e.g., $\alpha=\alpha_0$ and $D=D_0$, then \begin{equation}\label{constant}p(\alpha,D)=\delta(\alpha-\alpha_0)\delta(D-D_0) \,,\end{equation} 
where $\delta(\cdot)$ is the Dirac delta function.

Now we introduce the time-averaged mean-squared displacement (TAMSD)~\cite{Grebenkov2011b,Grebenkov2011} $M_{N}(\tau)$ computed from a trajectory as follows
\begin{equation}
    M_N(\tau)=\frac{1}{N-\tau}\sum_{n=1}^{N-\tau}(X(n+\tau)-X(n))^2,
\end{equation}
where $\tau=t/\Delta t$ and $\Delta t$ is the time elapsed between two consecutive position recordings.
It is known that, 
if the process is ergodic, its expectation equals the EAMSD, i.e.,
$\langle M_N(\tau)\rangle=2D(\tau\Delta t)^\alpha$. 
We mention 
that our method can also be formulated
in terms of other time-averaged statistics, see, e.g.,  \cite{MARAJ2020100017}. 
When applying a linear fitting of the log-log transformed TAMSD, the estimators of anomalous exponent and diffusion coefficient are given by 
\begin{equation}\label{eq:anomalous_exponent_estimator}
 \hat{\alpha}= \frac{\sum\limits_{\tau=1}^{\tau_{\max}}\ln(\tau)\ln
 \left(
 \frac{M_N(\tau)}{M_N(1)} \right)
}
{\sum\limits_{\tau=1}^{\tau_{\max}}\ln(\tau)^2} \,,
\end{equation}
\begin{equation}\label{eq:diffusion-coefficient_estimator}
 \hat{D}=\frac{1}{2}\exp\left(h\left(\mathbf{M}_N\right) -\hat{\alpha} B\right) \,,
\end{equation}
where
\begin{equation}
 h\left(\mathbf{M}_N\right)=
 \frac{1}{\tau_{\max}}\sum_{\tau=1}^{\tau_{\max}}\ln(M_N(\tau)) \,,
~~B=\frac{1}{\tau_{\max}}\sum_{\tau=1}^{\tau_{\max}}\ln(\tau\Delta t) \,.
\label{eq:hB}
\end{equation} 
The estimator given in Eq.~(\ref{eq:anomalous_exponent_estimator}) is purposefully a function of the ratios $M_N(\tau_i)/M_N(1)$ such that it is independent of both $D$ and $\Delta t$. In the above, $\tau_{\max}$ is the maximum lag time used in the fitting procedure. For conciseness, we use the notation $\mathbf{M}_N=\lbrace M_N(1),\ldots,M_N(\tau),\ldots,M_N(\tau_{\max})\rbrace$, which emphasizes the use of all the TAMSD values corresponding to the lag times $\tau=1,2,...,\tau_{\max}$.
The estimators in Eq.~(\ref{eq:anomalous_exponent_estimator}) and Eq.~(\ref{eq:diffusion-coefficient_estimator}) are mathematical objects that do not share the physical limitations of $\alpha$ and $D$. One can show that the anomalous exponent estimator has an upper bound by considering the ballistic case with a constant speed $v$, where $M_{N}(\tau)=2v\tau^2$. This serves as a physical upper bound for the TAMSD when there is no possible acceleration. Any random perturbation of this ideal case would slow down space exploration, thus lowering the anomalous exponent. Therefore, we conclude $\hat{\alpha}\leq 2$. In turn, for a perfectly immobile particle, the TAMSD is $M_N(\tau)=0$ from which the estimator diverges $\hat{\alpha}_{min}=-\infty$. 
For the estimated diffusion coefficient, it holds 
$\hat{D}\in[0,\infty)$ as expected.

We now proceed to derive the transfer function.
Let us consider the random variable $\hat{Y}=\ln(2\hat{D})$, based on Eq.~\ref{eq:diffusion-coefficient_estimator}
it reads
\begin{equation}
\label{eq:hat_Y_definition}
 \hat{Y}= h\left(\mathbf{M}_N\right)-\hat{\alpha} 
 B \,.
\end{equation}
First, we construct the intermediate transfer function $p(\hat{\alpha},\hat{Y}|\alpha,D)$ and then obtain the transfer function by changing variables.
It is well known that the joint PDF of two correlated random variables can be expressed through a bivariate Gaussian distribution under the condition of Gaussian marginal distributions and linear correlations. However, if either or both of these conditions are not fulfilled, this approach is not applicable. In this case, copula theory \cite{copula1} is the generalized framework that allows one to express the joint PDF (also applicable to dimensions higher than 2). To build the joint PDF in a two-dimensional version, Sklar's theorem \cite{copula2} tells us that three ingredients are needed: the marginal distributions of both variables and their correlation structure. There are multiple types of correlation structures, such as independent, linear (Gaussian copula), exponential, etc. 
In general, the marginal PDFs of $\hat{\alpha}$ and $\hat{Y}$ (i.e., the PDFs of univariate random variables $\hat{\alpha}$ and $\hat{Y}$, respectively)  are not Gaussian; therefore, copula modeling is justified in the considered case. However, given Eq.~(\ref{eq:hat_Y_definition}), $\hat{Y}$ and $\hat{\alpha}$ are linearly correlated; therefore, we chose a Gaussian copula density to model their correlation structure. Using this theory, one can show that the transfer function takes the form
\begin{eqnarray}\label{eq:joint_pdf_alpha_D_copula}
&&p(\hat{\alpha}, \hat{D} | \alpha,D)=\\\nonumber
&&c(Q_{\hat{\alpha}}(\hat{\alpha}),Q_{\hat{Y}}(\ln [2\hat{D}]),\rho)
\,
q_{\hat{\alpha}}(\hat{\alpha})\,
\frac{1}{\hat{D}}q_{\hat{Y}}(\ln[2\hat{D}]) \,,
\end{eqnarray}
where $c(u,v,\rho)$ is the Gaussian copula density with correlation coefficient $\rho$, marginal cumulative distribution functions (CDFs) $Q_{\hat{\alpha}}(\hat{\alpha})$ and $Q_{ \hat{Y} }( \hat{Y} )$ of $\hat{\alpha}$ and $ \hat{Y} $, respectively, and corresponding PDFs $q_{\hat{\alpha}}(\hat{\alpha})$ and $q_{ \hat{Y} }( \hat{Y} )$. For clarity, we omit 
for a while the conditional notation for the marginal PDFs, marginal CDFs (i.e., the CDFs of $\hat{\alpha}$ and $\hat{Y}$), and moments of the estimators, although they are all implicitly conditional on $\alpha$ and $D$.

The formula for the Gaussian copula density is
\begin{equation}\label{copula11}
\hspace{-0.3truecm} c(u,v,\rho)=\frac{1}{\sqrt{1-\rho^2}}\exp\left(-\frac{\rho^2 (x^2+y^2)-2\rho xy}{2(1-\rho^2)}\right) \,,
\end{equation}
where $x=\Phi^{-1}(u)$, $y=\Phi^{-1}(v)$, and $\Phi^{-1}(\cdot)$ denotes the inverse of standard Gaussian CDF.
In our case, the correlation coefficient $\rho$ in Eq.~(\ref{copula11}) is given by 
\begin{equation}\label{eq:corr_coeff_alpha_Y}
 \rho=\frac{ \langle \hat{Y} ^* \hat{\alpha}^*\rangle}
 {\std (\hat{Y})\, \std(\hat{\alpha})}
 \,,
\end{equation}
where $\hat{\alpha}^*=\hat{\alpha}-\langle\hat{\alpha}\rangle$, $\hat{Y}^*=\hat{Y}-\langle\hat{Y}\rangle$.


Importantly, this formulation allows us to explain why single particle experiments~\cite{cherstvy2019non,speckner2021single,benelli2021sub,korabel_etal-e-2021} consistently report the relationship $\hat{D}\propto\exp(\hat{\alpha} c_1+c_2)$. More precisely, $\hat{D}$ is plotted as a function of $\hat{\alpha}$ and an exponential trend is observed. The quantity of interest is the conditional expectation of $\hat{D}$ given $\hat{\alpha}$ noted $\langle\hat{D}|\hat{\alpha}\rangle$. We first notice from Eq.~(\ref{eq:hat_Y_definition}) that $\hat{\alpha}$ and $\hat{Y}$ are linearly dependent.
Additionally, when $\hat{Y}$ and $\hat{\alpha}$ are jointly Gaussian variables, the conditional expectation of $\hat{Y}$ verifies
\begin{equation}
\frac{\langle\hat{Y}|\hat{\alpha}\rangle-\langle\hat{Y}\rangle}{\std(\hat{Y})}
=\rho\,
\frac{\hat{\alpha}-\langle\hat{\alpha}\rangle}{\std(\hat{\alpha})},
\end{equation}
otherwise $\langle\hat{Y}|\hat{\alpha}\rangle$ can be expressed from copula theory~\cite{Crane2008} but is not necessarily proportional to $\hat{\alpha}$. 
The condition under which Gaussianity is valid is discussed later (see the discussion around Eq.(\ref{eq:condition_gaussianity})). In conjunction with the identity $\hat{Y}=\ln(2\hat{D})$ we get
\begin{equation}\label{eq:relation_D_alpha_exp}
 \langle\hat{D}|\hat{\alpha}\rangle=\frac{1}{2}\exp\left[\kappa\,(\hat{\alpha}-\langle\hat{\alpha}\rangle)+\langle\hat{Y}\rangle\right],
\end{equation}
where $\kappa=\langle \hat{\alpha}^* \hat{Y} ^*\rangle/ \var(\hat{\alpha})$.
The exponential behavior is a generic signature of performing a linear fitting in log-log scale, however, the parameters depend implicitly on $\alpha$ and $D$ thus reflecting the underlying physical process. This relation can be easily verified against experimental data.


As long as the anomalous exponent estimator is far enough from the upper bound, its marginal PDF is well approximated by a Gaussian distribution with mean $\langle \hat{\alpha}\rangle$ and standard deviation $\std(\hat{\alpha})$ as shown in Fig.~\ref{fig:beta_fit_exponent_nonoise_constant_D} (1.A-B).
Therefore, following the $3\sigma$ rule, we can set a simple criterion for the validity of the Gaussian approximation
\begin{equation}\label{eq:condition_gaussianity}
 \langle \hat{\alpha}\rangle +3\,\std(\hat{\alpha})<2 \,.
\end{equation}
 The aforementioned condition can be readily confirmed using the experimental data. Specifically, given the number of trajectories obtained from the experiment, we initially estimate the anomalous exponent for each trajectory individually. Subsequently, the expectation $\langle \hat{\alpha}\rangle$ and the standard deviation $\std(\hat{\alpha})$ in condition (\ref{eq:condition_gaussianity}) are substituted with their empirical equivalents derived from the data. Additionally, it is worth noting that $\std(\hat{\alpha})$  decreases with increasing trajectory length, therefore, the maximum admissible $\alpha$ for which the condition (\ref{eq:condition_gaussianity}) holds can be increased by taking longer trajectories (see SM~\ref{sec:SM}).

When condition (\ref{eq:condition_gaussianity}) is satisfied, then $\hat{Y}$ is also well approximated by a Gaussian distribution, see Fig.~\ref{fig:beta_fit_exponent_nonoise_constant_D} (2.A-C). When both marginal PDF of $\hat{\alpha}$ and $\hat{Y}$ are Gaussian, then the conditional joint PDF $p(\hat{\alpha}, \hat{Y} |\alpha,D)$ reduces to a bi-variate Gaussian distribution
\begin{equation}\label{eq:cond_joint_PDF_alpha_Y}
p(\hat{\alpha}, \hat{Y} |\alpha,D)
=\frac{1}{2\pi \sqrt{\det(C)}}
\exp\left(
 -\frac{1}{2}V_{\hat{\alpha}, \hat{Y} }^\top
 C^{-1}V_{\hat{\alpha}, \hat{Y} }
\right),
\end{equation}

where
\begin{equation} \label{eq:gaussian_joint_PDF_covariance_param}
 \begin{array}{ccc}
 V_{\hat{\alpha}, \hat{Y} }=
 \left(\begin{array}{c}
 \hat{\alpha}-\langle \hat{\alpha}\rangle\\
 \hat{Y} -\langle \hat{Y} \rangle 
 \end{array}\right),
 &
 C=\left[\begin{array}{cc}
 \var(\hat{\alpha}) & \langle \hat{\alpha}^* \hat{Y} ^*\rangle \\
\langle \hat{\alpha}^* \hat{Y} ^*\rangle & \var(\hat{Y})
 \end{array}
 \right].
 \end{array}
\end{equation}

From the above, after the change of variable 
$\hat{Y}=\ln(2\hat{D})$ we obtain
\begin{equation}\label{eq15}
p(\hat{\alpha}, \hat{D} |\alpha,D)
=\frac{1}{2\hat{D}\pi \sqrt{\det(C)}}
\exp\left(
 -\frac{1}{2}V_{\hat{\alpha}, \hat{D} }^\top
 C^{-1}V_{\hat{\alpha}, \hat{D} }
\right),
\end{equation}
where
\begin{equation}\label{eq:bivariate_cond_joint_pdf_VECTOR}
 V_{\hat{\alpha}, \hat{D} }=
 \left(\begin{array}{c}
 \hat{\alpha}-\langle\hat{\alpha}\rangle
\\
 \ln(2\hat{D}) -\langle \hat{Y} \rangle 
 \end{array}\right).
\end{equation}
The joint conditional PDF given in Eq.~(\ref{eq15}) is a simple, yet powerful approximation under condition (\ref{eq:condition_gaussianity}). Formulas (\ref{eq:corr_coeff_alpha_Y})-
(\ref{eq:bivariate_cond_joint_pdf_VECTOR}) can be computed based on the expectations of $\hat{\alpha}$ and $\hat{Y}$ as well as their variances and covariance for which explicit approximate expressions for arbitrary Gaussian processes are provided in the SM~\ref{sec:SM} (based on quadratic form statistics~\cite{Magnus1978}). To compute the moments, we assume that the standard deviation of TAMSD is small compared to its expectation. Therefore, the ergodicity breaking parameter~\cite{Deng2009} must be small. Conveniently, this parameter can be made as small as needed by increasing the length of trajectories.

From Eq.~(\ref{eq15}), by averaging over $\hat{\alpha}$, the marginal PDF of $\hat{D}$ conditioned on the true values $\alpha$ and $D$ has a log-normal distribution with moments described in SM~\ref{sec:SM}. Here, the log-normal distribution of $\hat{D}$ is solely due to the resolution limit resulting from the finite duration of the trajectories. Therefore, this distribution is distinct from the superstatistical log-normal distribution of the physical $D$ considered in reference~\cite{Beck2005}.

When condition~(\ref{eq:condition_gaussianity}) is not fulfilled, both marginal PDF of $\hat{\alpha}$ and $\hat{Y}$ becomes skewed. In such a case, other distributions may be more appropriate to describe the marginals of estimated parameters. In our analysis, we use the scaled-shifted Beta (called later Beta) distribution for $\hat{\alpha}$ as shown in Fig. ~\ref{fig:beta_fit_exponent_nonoise_constant_D} (1.A-C) and the Pearson IV distribution~\cite{Pearson1895,Nagahara1999} for $\hat{Y}$ as illustrated in Fig.~\ref{fig:beta_fit_exponent_nonoise_constant_D} (2.A-C)). In SM~\ref{sec:SM}, the main properties of both distributions are presented. 

 We check our analytical result with simulations of fBm. For clarity, we consider $p(\alpha,D)$ to be Dirac distributed (see Eq.~(\ref{constant})), however, through Eq.~(\ref{eq:joint_pdf_general_case}) the results hold for arbitrary $p(\alpha,D)$. In Fig.~\ref{fig:beta_fit_exponent_nonoise_constant_D} (columns (1) and (2)), we present the histograms of estimators $\hat{\alpha}$ (1.A-C) and $\hat{Y}$ (2.A-C) obtained from simulated $M=10^5$ fBm trajectories of $N=100$ steps with $\Delta t=1$ and $\tau_{\max}=5$. The anomalous exponent $\alpha=\lbrace 0.2,1,1.6\rbrace$ for rows A, B, and C, respectively. In each case, we assume $D=1$. Solid blue lines correspond to the theoretical curves based on the Gaussian approximations while the yellow lines correspond to the estimated Beta PDFs (in case of $\hat{\alpha}$) and Pearson IV PDF (in case of $\hat{Y}$). 
\begin{figure*}[!htbp]
\includegraphics[width=\textwidth]{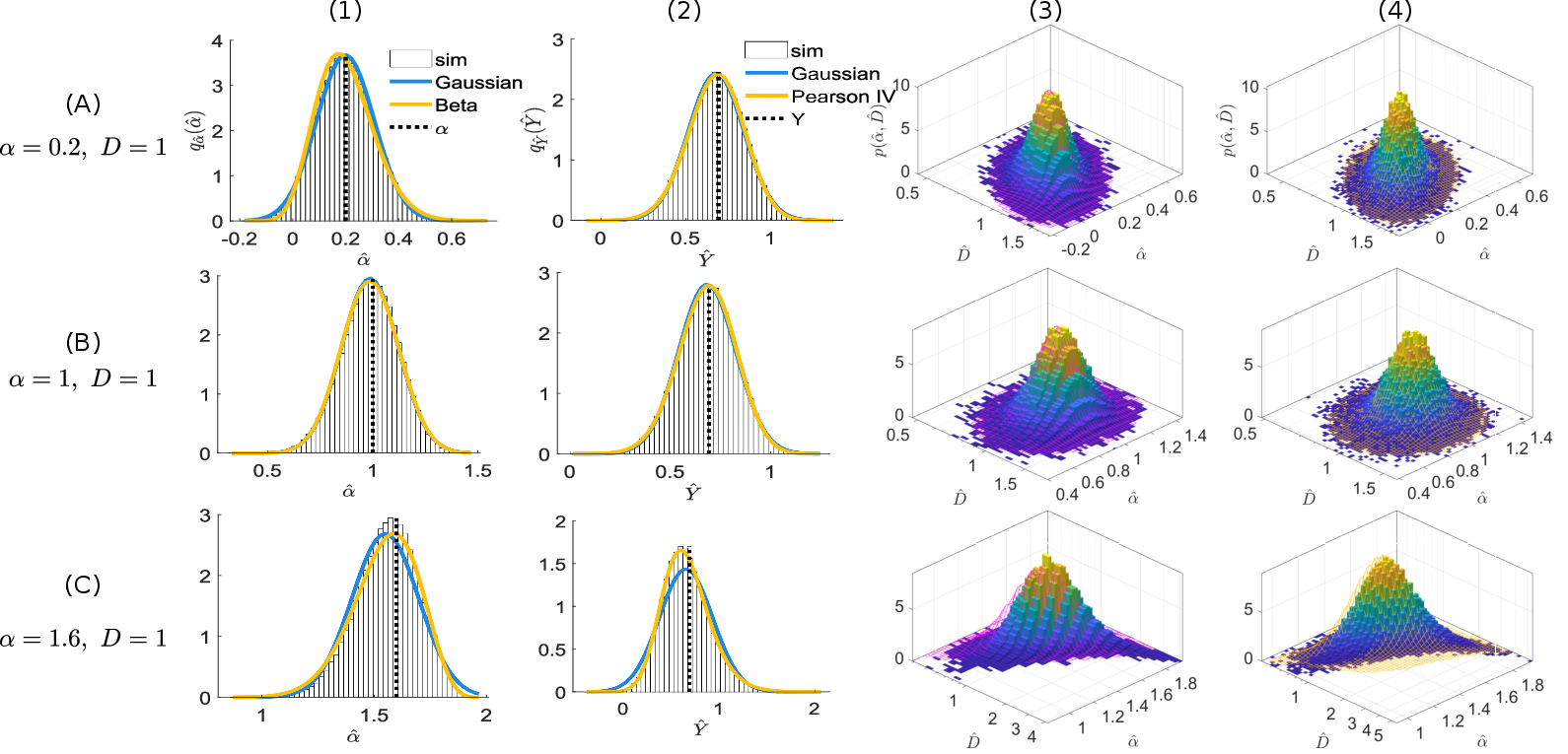}
 \caption{
(1): Histograms of $\hat{\alpha}$ versus theoretical predictions for Gaussian (blue line) and Beta distribution (yellow line); 
(2): Histograms of $\hat{Y}$ versus theoretical prediction for Gaussian (blue line) and Pearson IV (yellow line).
(3): histogram $p(\hat{\alpha},\hat{D})$ versus theory (magenta mesh) using Gaussian approximation of $\hat{\alpha}$ and $\hat{Y}$ from Eq.~(\ref{eq15}).
(4): histogram $p(\hat{\alpha},\hat{D})$ versus theory (yellow mesh) using copula-based approach Eq.~(\ref{eq:joint_pdf_alpha_D_copula}).
All the empirical histograms were obtained from $M=10^5$ simulated fBm trajectories of $N=100$ steps with $\tau_{\max}=5$ and $\alpha=\lbrace 0.2,1,1.6\rbrace$ for rows A, B, C. In each case, we choose $D=1$ and $\Delta t=1$. For the yellow lines and mesh columns (1), (2), (4), the skewness and kurtosis used in marginal PDFs were obtained empirically. 
}\label{fig:beta_fit_exponent_nonoise_constant_D}
\end{figure*}
In Fig.~\ref{fig:beta_fit_exponent_nonoise_constant_D} (3.A-C) we present the joint PDF $p(\hat{\alpha},\hat{D})$ estimated with Gaussian approximation of $\hat{\alpha}$ and $\hat{Y}$ distributions for the simulated trajectories of fBm (see the caption of the figure for more details). 
The theoretical joint PDFs coincide with the empirical ones for $\alpha<1.5$. The case $\alpha>1.5$ is trickier for applying Eq.~(\ref{eq:joint_pdf_alpha_D_copula}) to fBm. It has been shown~\cite{Deng2009} that in this regime the ratio of standard deviation over expectation of TAMSD for fBm decays as slowly as $1/N^{2-\alpha}$ such that much longer trajectories are required to get an ergodic behavior. In parallel, the correlation between $\hat{\alpha}$ and $\hat{Y}$ becomes non-linear for $\alpha>1.5$ near the tails of the transfer function. A more general copula may fit the non-linear correlation but it is out of the scope of the present article. In SM ~\ref{sec:SM} we present the maximum admissible $\alpha$ verifying condition~(\ref{eq:condition_gaussianity}) as a function of trajectory length.

In Fig.~\ref{fig:beta_fit_exponent_nonoise_constant_D} (4.A-C) we also present the joint PDF $p(\hat{\alpha},\hat{D})$ received using a copula-based approach. As can be seen, in such a case, the empirical and theoretical PDFs also coincide for large values of the anomalous exponent.
In Fig.~1 of SM~\ref{sec:SM}
a case is presented where the true $\alpha$ and $D$ are randomly distributed. The joint distribution function of $\alpha$ and $D$ is here represented as a discrete mixture
\begin{equation} \label{discrete}
 p(\alpha,D)
 =
 \sum_{k=1}^{m}
 c_k\,\delta(\alpha-\alpha_k)\delta(D-D_k),
\end{equation}
where $\sum_{k=1}^{m}
 c_k=1$.  The protocol for inferring discrete and continuous joint distributions $p(\alpha,D)$ is presented in the End Matter~\ref{sec:End_Matter}.

In addition, in SM~\ref{sec:SM} we demonstrate the influence of trajectory length $N$ on the accuracy of both the Gaussian approximation and the copula-based approach through the analysis of the corrected Hellinger distance based on \cite{hel}. The presented approaches are considered quite accurate when the distance is lower than 5\%. This is the case when $N\geq20$ for $\alpha\leq 1$ with Gaussian approximation.

In conclusion, we have presented the first complete characterization of the joint PDF of estimated anomalous diffusion parameters as a function of the underlying physical parameters. 
The proposed methodology is based on the copula theory for estimating the joint PDF and it 
performs well even for short trajectories ($\geq 30$ steps), which is critical for analyzing experimental data. It is known that reconstructing the joint distribution $p(\alpha, D)$ is crucial for correctly interpreting the dynamics and identifying changes in the physical parameters, particularly when following system perturbations (e.g., drug treatments in cellular biology). 
The conditional joint PDF can be characterized for any trajectory duration. Therefore, its knowledge enables a super-resolved joint PDF of the physical parameters beyond technical limitations due to trajectory duration.
The PDF depends on the moments of the quadratic form, for which we provided analytical expressions in the case of general Gaussian processes. This framework enables the separation of variability in the estimated physical parameters, distinguishing between the intrinsic randomness of the trajectories and the distributional behavior of the physical parameters themselves. 
Our approach is general and it paves the way to derive the PDF of any sufficiently smooth function of a quadratic form in Gaussian random variables. The practical aspect of this study is the development of universal estimators for anomalous diffusion parameters, which provide a reliable way to separate true parameter variability from measurement limitations in experiments. Thus, improving the analysis of molecular motion in heterogeneous media and enabling more accurate interpretations of data in various contexts. Future work will extend this approach to account for experimental errors, which further complicate the parameter estimation. Future work will also focus on applying the presented methodology to determine whether experimental data originate from a model with constant or random parameters, similar to the approach proposed in \cite{10.1063/5.0201436}. In this context, the Hellinger distance that measures the distance between the PDF of $(\hat{\alpha}, \hat{D})$ received for the experimental data and model with constant parameters may serve as an effective test statistic.
\textit{Acknowledgments:} GP and YL acknowledge the support by the Basque Government through the BERC 2022--2025 program and by the Ministry of Science and Innovation: BCAM Severo Ochoa accreditation CEX2021-001142-S / MICIN / AEI / 10.13039/501100011033. The work of AW was supported by the National Center of Science under Opus Grant 2020/37/B/HS4/00120. 

\textit{Code Availability:} The MATLAB code for fitting the joint distribution of estimated $\hat{\alpha}$ and $\hat{Y}$, as described in the End Matter~\ref{sec:End_Matter}, is available at \url{https://github.com/YannLanoiselee/Super-resolved-anomalous-diffusion-code}.
\clearpage
\title{End Matter}
\maketitle
\label{sec:End_Matter}
\label{sec:End_Matter_data_analysis}
In this section, we present how to use the proposed theory to analyze anomalous diffusion in an experimental setup. We have attached a MATLAB code to this manuscript that performs the fitting procedure based on $M$ trajectories, each with $N$ points, recorded with the same temporal resolution $dt$. The properly established conditional joint PDF $p(\hat{\alpha}, \hat{D} | \alpha, D)$ can be used to deduce the distribution of the true parameters $p(\alpha, D)$ based on the estimated joint PDF 
$p(\hat{\alpha}, \hat{D})$ by using Eq.~(\ref{eq:joint_pdf_general_case}). We study both the cases of a discrete and a continuous distribution of the true parameters $p(\alpha,D)$.

We consider here the case when the pair $\alpha,D$ might be randomly distributed. The joint PDF $p(\alpha, D)$ 
in Eq. (\ref{eq:joint_pdf_general_case}) is either a collection of Dirac delta functions (discrete mixture), or any two-dimensional PDF of random variables $\alpha$ and $D$. The minimal example, i.e., the discrete mixture given in Eq.~(\ref{discrete}), is explored here. 
Indeed, if a single component is sufficient to fit the joint PDF, one can conclude that there is a unique pair of 
$\alpha, D$. In such a case, the randomness of the estimated parameters can be fully attributed to estimation errors rather than to the intrinsic randomness of the true $\alpha$ and $D$. In turn, when more than one component is present, the fitting procedure allows for the characterization of the distribution of the true $\alpha, D$ with only a few parameters (3 per component).


Below we present the step-by-step procedure to fit an experimentally obtained joint PDF $p(\hat{\alpha},\hat{D})$ by assuming $p(\alpha,D)$ be a discrete mixture as stated in Eq.~(\ref{discrete}):\\
 (1). Select an ensemble of trajectories that are assumed to have fixed parameters along the trajectory. If there are changes in the dynamics over time, the trajectories should be segmented beforehand to obtain unchanging dynamics. Segmentation can be performed by using any relevant methodology~\cite{Lanoiselee2021,Stanislavsky2024,Qu2024}. (2). Select trajectory segments of the same number of steps, say $N^*\geq 20$. (3). Compute the TAMSD for each trajectory.
 (4). Estimate the anomalous exponent $\hat{\alpha}$ and diffusion coefficient $\hat{D}$ from the previously computed TAMSD by using Eq.~(\ref{eq:anomalous_exponent_estimator}) and Eq.~(\ref{eq:diffusion-coefficient_estimator}). (5). Make the transformation $\hat{Y}=\ln(2\hat{D})$. (6). Construct the $2$-dimensional empirical joint PDF $p(\hat{\alpha},\hat{Y})$ on the basis of the values of
 $\hat{\alpha}$ and $\hat{D}$. (7). Suppose that the underlying distribution of true $\alpha$ and $D$ is an $m$-components mixture. (8). Find the $m$ peaks of the empirical joint PDF $p(\hat{\alpha},\hat{Y})$. This is done in our script using \textit{fitgmdist} function from MATLAB \cite{gaussian_mixtures}. The location of the peaks are noted $\langle \hat{\alpha}_k^0\rangle$ and $\langle \hat{Y}_k^0\rangle$ where $k\in[1,m]$. After this,
estimate the respective weights $c_k^0$ of each peak. (9). Solve the inverse problem of deducing the true $\alpha_k^0$ and $D_k^0$ based on the expectations $\langle\hat{\alpha}_k^0\rangle$ and $\langle\hat{Y}_k^0\rangle$ by using the analytical formulas given in SM~\ref{sec:SM}. This gives a prior of the true parameters $\alpha_k,D_k$ of the $m$-component mixture of $p(\alpha,D)$ that will be refined in the next step. (10). Use the values $\alpha_k^0$, $D_k^0$ and $c_k^0$ as initial conditions for a least squares fitting that fits the empirical joint PDF $p(\hat{\alpha},\hat{Y})$ with formula Eq.~(\ref{eq:joint_pdf_general_case}), where the conditional joint distribution is given by Eq.~(\ref{eq:cond_joint_PDF_alpha_Y}) and the joint distribution of the true $\alpha$ and $D$ is assumed to be given by Eq.~(\ref{discrete}). There are only three parameters to be fitted per component 
 $\alpha_k$, $D_k$ and $c_k$. 

Note that, in general, the fitting could also be performed on $p(\hat{\alpha},\hat{D})$ instead of $p(\hat{\alpha},\hat{Y})$. 
However, since $\hat{Y}$ has a Gaussian distribution, it is more stable than the log-normally distributed $\hat{D}$.

The least squares fit, in principle, requires computing the moments of $(\hat{\alpha},\hat{Y})$ for each test value 
$(\alpha_k,D_k)$, which can be quite inefficient. 
To resolve this problem, we precompute the expectation, variance, and covariance values of $(\hat{\alpha},\hat{Y})$ for a reasonable ensemble of pairs $(\alpha,D)$, 
to build a 2-dimensional lattice of known values. Since all the statistics of $\hat{\alpha}$ and $\hat{Y}$ are smooth functions of $\alpha$ and $D$, one can deduce any expected statistics of $\hat{\alpha}$ and $\hat{Y}$ by interpolating the precomputed lattice. The interpolation must be nonlinear and smooth to prevent numerical errors in the gradient estimation during curve fitting.

In Fig.~5 of SM~\ref{sec:SM} we present an example of the curve fitting by using the code provided in the manuscript. 
It is based on the same example 
of a three-component mixture distribution of $p(\alpha,D)$ and with the same parameters as in Fig.~1 of SM~\ref{sec:SM}. However, here the estimation is just based on a thousand trajectories. 
This example highlights that the estimation is quite accurate for the three components. Additionally, it shows that fitting only the histogram of $\hat{\alpha}$ or that of $\hat{D}$ can be misleading when two or more components have the same $\alpha$ or $D$. Only a fitting of the joint PDF of the estimated parameters leads to a correct assessment of the components of $p(\alpha,D)$.

Another highly relevant problem is when $p(\alpha,D)$ assumes a continuous joint distribution. The shape of the joint distribution $p(\alpha,D)$ is not universal, it depends on the system considered. However, physical constraints for diffusive systems tell us that $D\geq 0$ and $\alpha\in(0,2]$. To comply with the physical limitations, we assume that $\alpha$ has a Beta distribution on the interval $\alpha\in(0,2)$,
and that $D$ has a log-normal distribution such that $D\in(0,\infty)$. For simplicity, we assume linear correlation with coefficient $\rho_{\alpha,D}$. The joint distribution of the true parameters $p(\alpha,D)$, can be expressed using the copula theory
\begin{equation}\label{eq:joint_PDF_true_alpha_D}
    p(\alpha,D)=
    c(Q_{\alpha}(\alpha),Q_{D}(D)),\rho_{\alpha,D})
\,
q_{\alpha}(\alpha)\,
q_{D}(D) \,,
\end{equation}
where $c(u,v,\rho)$ is the Gaussian copula PDF defined in Eq.~(\ref{copula11}), $q_{\alpha}(\alpha)$, $q_{D}(D)$ are the marginal PDFs of physical $\alpha$ and $D$, and $Q_{\alpha}(\alpha)$, $Q_{D}(D)$ are the corresponding marginal CDFs.
The fitting strategy from the previous section can be applied here up to point 6. Under the assumption that $p(\alpha,D)$ is distributed according to Eq.~(\ref{eq:joint_PDF_true_alpha_D}), a least square fitting is performed over the joint PDF of estimated parameters $p(\hat{\alpha},\hat{Y})$. 

In Fig.~\ref{fig:continijous_joint_fit} we present the results of the fitting for $M=10^3$ trajectories of duration $N=100$. We selected the parameters to ensure $\langle\alpha\rangle=1.4$, $\std(\alpha)=0.2$, $\langle D\rangle=1$ and $\std(D)=0.5$, and $\rho_{\alpha,D}=-0.7$. 
\begin{figure*}
\includegraphics[width=0.9\textwidth]{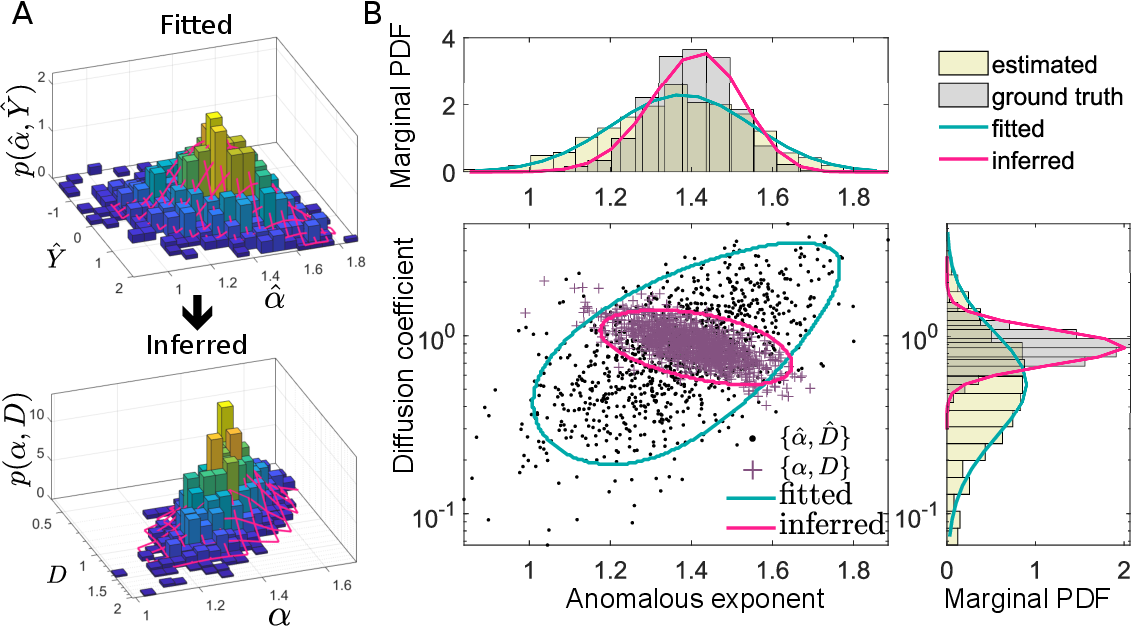}
 \caption{(A) Top: Histogram of the empirical $p(\hat{\alpha},\hat{Y})$ overlayed with the fitted distribution
 (A) Bottom: Histogram of the ground truth $p(\alpha,D)$ overlayed with the joint PDF inferred from the fitting of empirical $p(\hat{\alpha},\hat{Y})$.
 (B) Scatter plot of $\lbrace \hat{\alpha},\hat{D}\rbrace$ (black dots) and
 $\lbrace \alpha,D\rbrace$ (purple crosses) overlayed with teal isoline for the fitted joint PDF $p(\hat{\alpha},\hat{D})$
and pink isoline for the inferred joint PDF $p(\alpha,D)$. On top, histogram of ground truth $\alpha$ (grey bars) and $\hat{\alpha}$ (yellow bars), teal line is $p(\hat{\alpha})$ obtained from fitting and pink line is $p(\alpha)$ inferred from the fitting.
On the right, histogram of ground truth $D$ (grey bars) and $\hat{D}$ (yellow bars), teal line is $p(\hat{D})$ obtained from fitting and pink line is $p(D)$ inferred from the fitting.
 The figure is based on a single simulated data set of $M=10^3$ trajectories of fBm of $N=100$ steps with jointly randomly distributed $\alpha$ and $D$ according to Eq.~(\ref{eq:joint_PDF_true_alpha_D}) with parameters $\langle\alpha\rangle=1.4$, $\std(\alpha)=0.2$, $\langle D\rangle=1$, $\std(D)=0.5$ and $\rho_{\alpha,D}=-0.7$.}
\label{fig:continijous_joint_fit}
 \end{figure*}
The top histogram of Fig.~\ref{fig:continijous_joint_fit}.A  corresponds to the joint distribution of estimated parameters $p(\hat{\alpha},\hat{Y})$ together with the fitted distribution (mesh). The bottom histogram of Fig.~\ref{fig:continijous_joint_fit}. A corresponds to $p(\alpha,D)$ used in the simulation (inaccessible from experiments) overlayed with a mesh for the joint PDF of the true parameters inferred from the fitted $p(\hat{\alpha},\hat{Y})$.  
Fig.~\ref{fig:continijous_joint_fit}.B summarizes the results, the scatter plot shows the empirically estimated parameter pairs $\lbrace\hat{\alpha},\hat{D}\rbrace$ (black dots), the ground truth physical parameters 
$\lbrace\alpha,D\rbrace$ (purple crosses), a teal isoline containing most of the fitted joint PDF of estimated parameters lies and a pink isoline containing most of the joint PDF of the true parameters inferred from fitting the estimated parameters.
The top graph shows the marginal PDF of the anomalous exponent, yellow bars represent empirically estimated values and grey bars are the true values, teal line shows the marginal of the joint PDF fitted in Fig.~\ref{fig:continijous_joint_fit}.A and pink line is the marginal PDF obtained from the inferred marginal PDF of the true anomalous exponent. On the right, the same plots are shown but for the diffusion coefficients. The joint PDF of the true anomalous exponent and diffusion coefficient is well reconstructed, the inferences give $\langle\alpha\rangle=1.427$, $\std(\alpha)=0.094$, $\langle D\rangle=1.0075$, $\std(D)=0.487$. It is worth noting that measuring the Pearson correlation between the estimated parameters yields $\rho_{\hat{\alpha},\hat{D}}\approx 0.8$ which is a very different result compared to the correlation between the true ones ($\rho_{\alpha,D}=-0.7$). The correlation between the true parameters $\rho_{\alpha,D}$ is apparently destroyed during the estimation procedure, however it can be inferred using the procedure proposed here from which we obtained $\rho_{\alpha,D}=-0.816$.

To conclude this part, it is worth mentioning that the methodology presented in this article could form the basis for distinguishing between a continuous and a discrete two-dimensional random variable $(\alpha,D)$. We are not detailing such a procedure now, as it requires more in-depth analysis and verification for various scenarios. Therefore, we are merely addressing this problem, leaving it open for future research. However, our preliminary findings suggest that the type of distribution (discrete or continuous) of $(\alpha,D)$ can be distinguished with the present methodology. 


\clearpage
\title{Supplementary Materials}
\maketitle

\label{sec:SM}

\section{Example of the joint PDF of estimated parameters from randomly distributed true 
$\alpha$ and $D$}
In Fig.~1, a three-component mixture (see Eq.~(19) in the main text) is simulated with three pairs of anomalous parameters: ${\alpha_1=1,D_1=0.1}$, ${\alpha_2=1,D_2=0.15}$, and ${\alpha_3=0.1,D_3=0.1}$, with $c_1=c_2=c_3=1/3$. The parameters have been chosen such that, among the three components, two components have the same $\alpha$ and two components have the same $D$. The same $\alpha$ but different $D$ could be physically interpreted as a molecule with two distinct conformations. In turn, the same diffusion coefficient but different $\alpha$ could be interpreted as a molecule exploring a homogeneous part of the medium (Brownian motion with $\alpha=1$) or a more crowded part of the medium (subdiffusion with $\alpha<1$). In this example, the theoretical PDF based either on the copula approach or on the Gaussian approximation yields the same result because condition~(14) of the main text is fulfilled.
\begin{figure}[!h]
\includegraphics[width=0.5\textwidth]{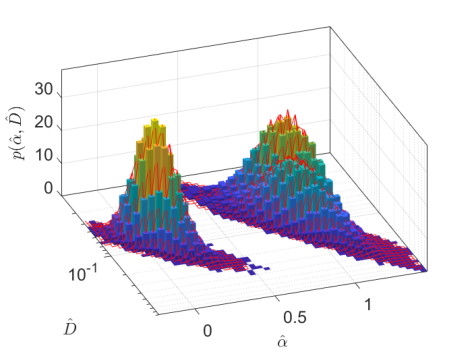}
 \caption{Histogram $p(\hat{\alpha},\hat{D})$ versus theory (magenta mesh) using Gaussian approximation of $\hat{\alpha}$ and $\hat{Y}$ from Eq.~(17) from main text where $p(\alpha,D)$ is distributed according to Eq.~(19) from main text with three components $\alpha_1=0.1$, $D_1=0.1$, $\alpha_2=1$, $D_2=0.1$, $\alpha_3=1$, $D_3=0.15$ with probability $c_1=c_2=c_3=1/3$. Empirical histogram was obtained from $M=10^5$ trajectories of $N=100$ steps with
 $dt=0.03$ and fitted using $\tau_{\max}=5$.
}\label{fig:joint_pdf_mixture_emp_vs_theory}
\end{figure}

\section{Statistical properties of functions of TAMSD}
\label{Annex:sec:statistical_propetis_estimator}
\label{sec:TAMSD_function_mean_covariance}
In this section, 
we establish the general formulas for the mean and the variance of a function of 
the time-averaged mean-squared displacement (TAMSD)
$\mathbf{M}_N$. These general results will then be applied to the case of the estimators 
$\hat{\alpha}$ and $\hat{Y}$. As such, all necessary moments are provided to compute analytically the conditional joint PDF $p(\hat{\alpha},\hat{D}|\alpha,D)$ in Eq.~(16) of the main text.  

\subsection{Expectation of a function of TAMSD}
Our goal is to find the expectation of  $f(\mathbf{M}_N)$ for a sufficiently smooth function $f(\cdot)$. This will allow us later to obtain statistics of the estimator of anomalous exponent and diffusion coefficient. 

First, we define the centered TAMSD 
\begin{equation}
M_N^*(\tau_i)=M_N(\tau_i)-\langle M_N(\tau_i)\rangle.
\end{equation}
Next, we proceed to a Taylor approximation around the means $\mathbf{\langle\mathbf{M}_N\rangle}$ up to the second order. The multidimensional Taylor expansion in matrix form reads 
\begin{eqnarray}\label{eq:Taylor_Expansion_function_of_TAMSD}
   f\left(\mathbf{M}_N\right) 
   &\approx&
f\left(\langle\mathbf{M}_N\rangle\right)
\\\nonumber 
&&+{\mathbf{M}_N^*}^\top \mathcal{D}f(\langle\mathbf{M}_N\rangle)
\\\nonumber 
 &&+
\frac{1}{2}{\mathbf{M}_N^*}^\top \mathcal{D}^2f(\langle\mathbf{M}_N\rangle)\mathbf{M}_N^*\\\nonumber 
 &&+\ldots,
\end{eqnarray}
where $\mathcal{D}f$ and $\mathcal{D}^2f$ are the gradient and the Hessian matrices, respectively, of $f(\cdot)$ with respect to the vector $\mathbf{M}_N$. Such expansion is valid when the standard deviation of TAMSD is small compared to its expectation. As the standard deviation decreases with increasing trajectory length $N$, there is always an $N$ for which the expansion can be performed accurately. 
Therefore, we obtain the mean of the function $f(\cdot)$
\begin{eqnarray}\label{eq:expectation_fun_quad_form}
    &&\langle f\left(\mathbf{M}_N\right) \rangle
    \\\nonumber
    &=&
    f\left(\langle\mathbf{M}_N\rangle\right)
+
\frac{1}{2}\langle {\mathbf{M}_N^*}^\top \mathcal{D}^2f(\langle\mathbf{M}_N\rangle)\mathbf{M}_N^* \rangle
 \\\nonumber
    &=&
  f\left(\langle\mathbf{M}_N\rangle\right)
+\frac{1}{2}\sum_{i,j=1}^{\tau_{max}}\partial_{ij}  f\left(\langle\mathbf{M}_N\rangle\right)\langle (M_N^{*}(\tau_i)M_N^{*}(\tau_j)\rangle.
\end{eqnarray}
Thus, it is clear that the mean of the function $f(\cdot)$ is affected by the covariance of TAMSD.
\\

\subsection{Covariance of two functions of TAMSD}

Concerning the covariance calculation, for simplicity, we abandon the quadratic term in Eq.~(\ref{eq:Taylor_Expansion_function_of_TAMSD}) to keep the linear approximation. The covariance of two functions $f(\cdot)$ and $g(\cdot)$ is
\begin{eqnarray}\nonumber
   &&\langle f^*\left(\mathbf{M}_N\right) g^*\left(\mathbf{M}_N\right)\rangle\\
   && \qquad =
\left\langle \left({\mathbf{M}_N^*}^\top \mathcal{D}f(\langle\mathbf{M}_N\rangle)\right)\left({\mathbf{M}_N^*}^\top \mathcal{D}g(\langle\mathbf{M}_N\rangle)\right)\right\rangle,
\end{eqnarray}
where the gradient is a diagonal matrix and $f^*\left(\mathbf{M}_N\right)=f\left(\mathbf{M}_N\right)-\langle f\left(\mathbf{M}_N\right)\rangle$. We deduce
\begin{eqnarray}\label{eq:covar_estimator_general}
   && \langle f^*\left(\mathbf{M}_N\right) g^*\left(\mathbf{M}_N\right)\rangle
   \nonumber \\
  && \quad =
\sum_{i,k}\partial_i f(\langle \mathbf{M}_N\rangle)\,
\langle M_N^*(\tau_i)M_N^*(\tau_k)\rangle\,\partial_k g(\langle \mathbf{M}_N\rangle),
\end{eqnarray} 
which can easily be computed as a quadratic form $ \langle f^*\left(\mathbf{M}_N\right) g^*\left(\mathbf{M}_N\right)\rangle=u^\top \Lambda v$ with vector elements $u_i=\mathcal{D}_i f(\langle \mathbf{M}_N\rangle)$ and $v_i=\mathcal{D}_i g(\langle \mathbf{M}_N\rangle)$ and the covariance matrix of TAMSD with elements $\Lambda_{ij}=\langle M_N^*(\tau_i)M_N^*(\tau_k)\rangle$.
Applying this to the anomalous exponent estimator from
Eq.~(5) in the main text and the other to $\hat{Y}$ from Eq.~(8) in the main text, we obtain the covariance $\langle \hat{\alpha}^*\hat{Y}^*\rangle$ in Eq.~(11).

The variance of a function $f(\cdot)$ of TAMSD can be computed in a very similar way
\begin{eqnarray}\label{eq:var_estimator_general}
   && \var (f\left(\mathbf{M}_N\right))
   \nonumber \\
  && \quad =
\sum_{i,k}\partial_i f(\langle \mathbf{M}_N\rangle)\,
\langle M_N^*(i)M_N^*(k)\rangle\,\partial_k f(\langle \mathbf{M}_N\rangle) .
\end{eqnarray}


 \section{Statistical properties of the estimators $\hat{\alpha}$ and $\hat{Y}$}
In the previous section, we presented the computations of the expectation and covariance of a function of TAMSD, which are crucial for computing the moments of $\hat{\alpha}$ and $\hat{Y}$.
Let us define
\begin{equation}\label{eq:anomalous_exponent_estimator}
f\left(\mathbf{M}_N\right)=\frac{\sum_{i=1}^{\tau_{\max}}\ln(\tau_i)\ln
 \left(
 \frac{M_N(\tau_i)}{M_N(1)} \right)
}
{\sum_{i=1}^{n}\ln(\tau_i)^2} \,,
\end{equation}
\begin{equation}\label{eq:diffusion-coefficient_estimator}
k\left(\mathbf{M}_N\right)=h\left(\mathbf{M}_N\right) -\hat{\alpha} B \,,
\end{equation}
\begin{equation}\label{eq:diffusion-coefficient_estimator}
 g\left(\mathbf{M}_N\right)=\frac{1}{2}\exp\left(k\left(\mathbf{M}_N\right)\right) \,,
\end{equation}
where $h\left(\mathbf{M}_N\right)$ and $B$ are defined in Eq.~(7) of the main text respectively.
which expresses the estimators of $\hat{\alpha}$, $\hat{Y}$ and $\hat{D}$ as functions of all values of TAMSD used in the fitting.

Applying Eq.~(\ref{eq:expectation_fun_quad_form}) to the estimators of $\hat{\alpha}$ and $\hat{Y}$ we obtain their expectations 
\begin{equation}\label{eq:expectation_alpha_Y}
 \left\lbrace
 \begin{array}{l}
\langle\hat{\alpha}\rangle\approx\alpha+\frac{1}{2}Tr[\Lambda^\top\mathcal{D}^2f(\langle\mathbf{M}_N\rangle)],
\\ 
\\ \langle\hat{Y}\rangle\approx Y
+\frac{1}{2}Tr[\Lambda^\top\mathcal{D}^2k(\langle\mathbf{M}_N\rangle)],
 \end{array}
 \right.
\end{equation}
where $\mathcal{D}^2f(\cdot)$ (resp. $\mathcal{D}^2k(\cdot)$) denotes the Hessian matrix of the function $f(\cdot)$ (resp. $k(\cdot)$) defined in Eq.~(4) (resp. Eq.~(8)) of the main text with respect to the variables $M_N(1),\ldots,M_N(\tau_{\max})$. 
Moreover, $\Lambda$ is the covariance matrix of $\mathbf{M}_N$, see Eq.~(\ref{eq:covar_TAMSD_fBm_noise_supp}). 
It should be noted that the expectations of the estimators yield the true values of $\alpha$ and $Y$ with corrections. The corrections depend on the covariance of TAMSD. When $N\to\infty$ but $\tau_{\max}$ is fixed, the distribution of TAMSD converges to a Dirac delta function $\delta(x-\langle M_N(\tau)\rangle)$ and the covariance of TAMSD vanishes. Thus, the estimators $\hat{\alpha}$ and $\hat{Y}$ are asymptotically unbiased. However, at finite $N$ the corrections given in Eq.~(\ref{eq:expectation_alpha_Y}) should be taken into account.
Applying Eq.~(\ref{eq:covar_estimator_general}) to the functions $f(\cdot)$ and $k(\cdot)$ yields the approximate covariance and variances
\begin{equation}\label{eq:variances_covariance_alpha_Y}
\left\lbrace
 \begin{array}{l}
 \var(\hat{\alpha})\approx
\mathcal{D}f(\langle \mathbf{M}_N\rangle)^\top\, \Lambda \,\mathcal{D}f(\langle \mathbf{M}_N\rangle), 
 \\
 \var(\hat{Y}) \approx \mathcal{D}k(\langle \mathbf{M}_N\rangle)^\top\, \Lambda \,\mathcal{D}k(\langle \mathbf{M}_N\rangle), \\
 \langle \hat{Y} ^* \hat{\alpha}^*\rangle \approx
 \mathcal{D}f(\langle \mathbf{M}_N\rangle)^\top\,\Lambda \,\mathcal{D}h(\langle \mathbf{M}_N\rangle)
 - \var(\hat{\alpha}) B.
 \end{array}
\right.
\end{equation}

The moments of estimators $\hat{\alpha}$ and $\hat{D}$ depend on the expectation $\langle M_N(\tau)\rangle$, covariance matrix $\Lambda$ of $M_N(\tau)$ which expressions we obtain exactly.\\
In general, the TAMSD can be computed as a quadratic form
\begin{equation}
    M_N(\tau)=\mathbb{Y}_N^\top R_\tau \mathbb{Y}_N,
\end{equation}
where $R_\tau$ is a matrix introduced in~\cite{Grebenkov2011} whose entries are \begin{equation}\label{eq:Rmatrix}
 R_\tau(n_1,n_2)=\frac{1}{N-\tau}(m_{n_1}\delta_{n_1,n_2}-\delta_{|n_1-n_2|,\tau}),
\end{equation}
where $\delta_{\cdot,\cdot}$ is the Kronecker delta, and 
\begin{equation}
 m_k=\left\lbrace \begin{array}{ll}
 \left\lbrace
 \begin{array}{cl}
 1 & k\leq \tau \textrm{ or } k> N-\tau \\
 2 & \tau<k\leq N-\tau 
 \end{array}
 \right.
 & \tau\leq N/2 \\
\left\lbrace
\begin{array}{cl}
 1 & k\leq N-\tau \textrm{ or } k> \tau \\
 0 &N-\tau<k\leq \tau
\end{array}
\right. & \tau>N/2. 
 \end{array}
 \right.
\end{equation}
Note that, compared to the original version from~\cite{Grebenkov2011}, here the matrix $R_\tau$ is divided by $2$ and we corrected typos in the inequality signs.\\
This allows us to compute the moments and covariance of $M_N$ using the theory of quadratic forms~\cite{Magnus1978}.
We note $\Sigma$ the covariance matrix of a Gaussian process. 
In this case, we have
\begin{equation}\label{eq:mean_TAMSD_with_noise_supp}
\langle M_N(\tau)\rangle=Tr[R_\tau\Sigma] \,.
\end{equation}
The covariance can also be computed explicitly 
\begin{eqnarray}\label{eq:covar_TAMSD_fBm_noise_supp}
    \Lambda_{ij}=\langle M_N^*(i)M_N^*(j)\rangle
      =
       2Tr[\Sigma R_i\Sigma R_j] \,.
   \end{eqnarray}
 This concludes the analytical computation by which, provided a Gaussian process with an arbitrary covariance matrix $\Sigma$, the moments of the estimator can be analytically approximated.

\section{Statistical properties of $\hat{D}$}
As discussed in the main text, $\hat{D}$ has a log-normal distribution.
Its moments can be expressed as the function of the moments of $\hat{Y}$. 
The expectation is
\begin{equation}
     \langle \hat{D} \rangle =\frac{1}{2}\exp\left(\langle\hat{Y}\rangle+\frac{1}{2}\var(\hat{Y})\right),
\end{equation}
and the variance is
\begin{equation}
     \var( \hat{D} ) =\frac{1}{4}\left(\exp\left(\var(\hat{Y})\right)-1\right)\exp\left(2\langle\hat{Y}\rangle+\var(\hat{Y})\right).
\end{equation}
Alternatively, the moments can be obtained from the Taylor expansion of the diffusion coefficient estimator, from which the expectation reads
\begin{equation}
 \langle \hat{D} \rangle
 \approx D
+\frac{1}{2}Tr[\Lambda^\top\mathcal{D}^2g(\langle\mathbf{M}_N\rangle)],
\end{equation}
where the function $g(\cdot)$ is defined in Eq.~(\ref{eq:diffusion-coefficient_estimator}).
The first-order approximation of the variance of $\hat{D}$ is as follows 
\begin{equation}\label{eq:var_estimator_general_D}
 \var (\hat{D})
\approx\mathcal{D}g(\langle \mathbf{M}_N\rangle)^\top\, \Lambda \,\mathcal{D} g(\langle \mathbf{M}_N\rangle).
\end{equation}

\section{Super-diffusive case}
\begin{figure}[!htbp]
\includegraphics[width=0.5\textwidth]{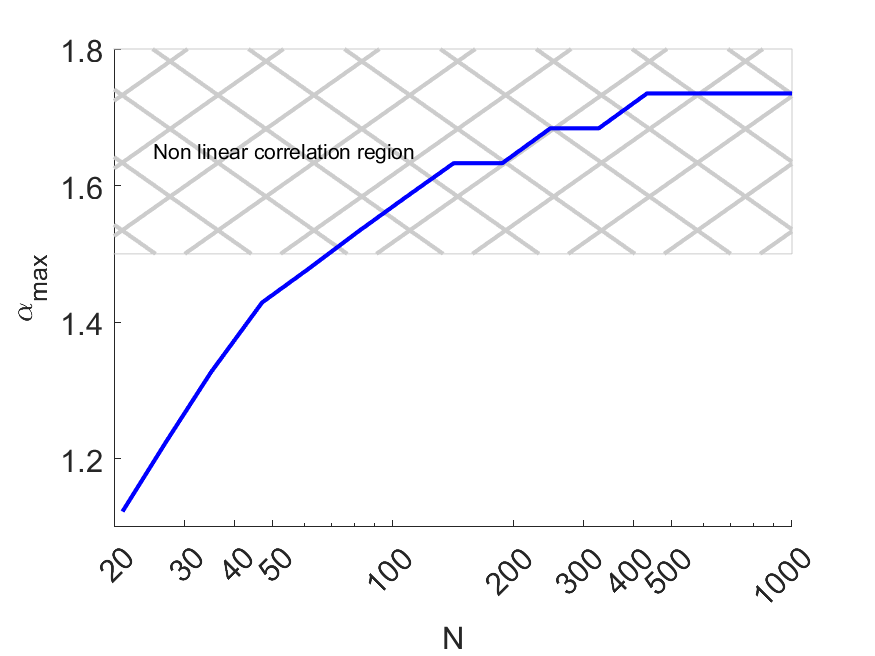}
 \caption{Theoretical curve of the maximum anomalous exponent $\alpha_{max}$ for which the Gaussian approximation holds as a function of trajectory length $N$ given a maximum lag-time $\tau_{max}=5$ for the fitting of TAMSD applied to fractional Brownian motion. \label{fig:alpha_max_fct_N}}
 \end{figure}

The super-diffusive case is slightly more complex, as the Gaussian assumption does not necessarily hold.
In Fig.~\ref{fig:alpha_max_fct_N}, we present the theoretical bound for the maximum true value of $\alpha$ for which the Gaussian assumption of both $\hat{\alpha}$ and $\hat{Y}$ is valid as a function of the trajectory length. This bound is valid for any $D$. As can be seen, most of the super-diffusive range is accessible but requires progressively longer trajectories.
This is because the Taylor approximation of the estimator around the expectations of the TAMSD does hold when TAMSD variance is small compared to its expectation. The ergodicity-breaking parameter $EB(N)$ measures the variance compared to the squared expectation of TAMSD. For fractional Brownian motion (fBm), it has been shown~\cite{Deng2009} that when $\alpha>1.5$ there is a switch from a $EB(N)\propto N^{-1}$ decay to a $EB(N)\propto N^{2(2-\alpha)}$. Unfortunately, for $\alpha>1.5$, the correlation between $\hat{\alpha}$ and $\hat{Y}$ becomes also non-linear therefore the present linear correlation approach is progressively invalid as $\alpha\to 2$, in turn applying different type of correlation structure (different copula) could work in this region as well. 

\section{Marginal distributions of $\hat{\alpha}$ and $\hat{Y}$ in the non-Gaussian case}
\subsection{Scaled-shifted Beta distribution for $\hat{\alpha}$}\label{Annex:sec:beta_pdf_for_exponent_pdf}
The scaled-shifted Beta distribution is a well-suited PDF to model the anomalous exponent estimator. It is a bounded distribution, therefore, it can account for the upper bound $\hat{\alpha}<2$. Additionally, one of its limit cases is the Gaussian distribution, so it is compatible with the approximation of the anomalous exponent estimator when the condition in Eq.~(14) of the main text is verified.

Suppose that $\hat{\alpha}$ is bounded such that $\hat{\alpha}_{min}\leq\hat{\alpha}\leq\hat{\alpha}_{max}$. The scaled-shifted Beta distribution is 
\begin{equation}
    p(\hat{\alpha})=\frac{(\hat{\alpha}-\hat{\alpha}_{min})^{a-1}(\hat{\alpha}_{max}-\hat{\alpha})^{b-1}}{B(a,b)(\hat{\alpha}_{max}-\hat{\alpha}_{min})^{a+b-1}}.
\end{equation}
An interesting parameterization for Beta distribution is based on its mean and variance by setting $a=\phi\mu$ and $b = \phi(1-\mu)$, where 
\begin{equation}
    \mu=\frac{\langle \hat{\alpha}\rangle-\hat{\alpha}_{min}}{\hat{\alpha}_{max}-\hat{\alpha}_{min}},
\end{equation}
and
\begin{equation}
    \phi=(\hat{\alpha}_{max}-\hat{\alpha}_{min})^2\frac{\mu(1-\mu)}{\var(\hat{\alpha})}-1 \,.
\end{equation}
The main interest of this parametrization is that $a$ and $b$ depend solely on the mean and variance of $\hat{\alpha}$. The bounds of the distribution $\alpha_{min}$ and $\alpha_{max}$ contain information about the skewness and kurtosis.

Because the estimator is not bounded on the interval $[0,2]$, one can use $\alpha_{min}\neq 0$ and $\alpha_{max}\neq 2$ to accommodate for the $\hat{\alpha}$ distribution. 
There are two options for determining the bounds. They can be either empirically estimated using
\begin{equation}
\left\lbrace 
\begin{array}{l}
         \hat{\alpha}_{min}=min([0,min(\hat{\alpha})]),  \\
           \hat{\alpha}_{max}=2
    \end{array}
\right.
\end{equation}
or, they can be theoretically computed using the method of moments, which requires knowledge of the first 4 moments.
\subsection{Pearson type IV distribution for $\hat{Y}$}
\label{Annex:sec:Pearson_pdf_for_Y_pdf}

The Pearson type IV distribution is a particular solution of the Pearson differential equation~\cite{Pearson1895}
\begin{equation}
    \frac{p(x)'}{p(x)}+\frac{a+(x-\mu)}{b_2(x-\mu)^2+b_1(x-\mu)+b_0}=0,
\end{equation}
of which particular solutions contain a large range of distributions as the Student's t-distribution, Beta distribution, Gamma distribution, etc.
The Pearson type IV distribution arose as a solution of this equation, it is less known than its other counterparts and is characterized by a high excess kurtosis and low skewness. 
For $b_1^2-4b_2b_0<0$, after making the change of variable $y=x+\frac{b_1}{2b_2}$, the solution is
\begin{equation}
    p(y)\propto \left(1+\frac{y^2}{\alpha^2}\right)^{-m}\exp\left(-\nu\,\textrm{arctan}\left(\frac{y}{\alpha}\right)\right),
\end{equation}
where $m=1/(2b_2)$, $\nu=-\frac{2b_2a-b_1}{2b_2^2\alpha}$, and $\alpha=\frac{\sqrt{4b_2b_0-b_1^2}}{2b_2}$. The parameters can be fitted using a maximum likelihood approach \cite{Nagahara1999} or by moment matching method using the expectation, variance, skewness, and kurtosis. Note that the Pearson type IV distribution converges to a Gaussian when both $m$ and $\nu$ are large.

\section{Precision of analytical approach as a function of trajectory length}
\subsection{Corrected Hellinger distance}
The Hellinger distance~\cite{hel}, a measure of divergence (statistical distance), quantifies the difference between two probability distributions (in quadratic norm). For two probability distributions with PDFs $d_1(\cdot)$ and $d_2(\cdot)$, it is defined as follows
\begin{eqnarray}\label{Hellinger}
 H(d_1,d_2)=\left(\int\left(\sqrt{d_1(x)}-\sqrt{d_2(x)}\right)^2dx\right)^{1/2}.
\end{eqnarray}
In practice, integral~(\ref{Hellinger}) is replaced by its empirical counterpart. In our study, the Hellinger distance is calculated in a two-dimensional context, where the densities $d_1(\cdot)$ and $d_2(\cdot)$ are the theoretical joint PDF $p(\hat{\alpha},\hat{D})$ and the empirical joint PDF of $\hat{\alpha}$ and $\hat{D}$, respectively.
When using the empirical counterpart, additional distance is introduced. More precisely, we define $d_{emp}$ to be the empirical joint PDF obtained from the data while $d_{th}$ be the theoretical joint PDF. The empirical Hellinger distance is then given by
\begin{eqnarray}\label{Hellinger1}
H(d_{emp},d_{th})=\left(\int\left(\sqrt{d_{emp}(x)}-\sqrt{d_{th}(x)}\right)^2dx\right)^{1/2}.
\end{eqnarray}
The Hellinger distance is zero when the two PDFs are exactly the same. However, the number of empirical trajectories $M$ used to build the empirical histogram $d_{emp}$ is finite. As a result, the height of the bins in $d_{emp}$ displays some degree of randomness. Therefore, the Hellinger distance is always larger than zero $H(d_{emp},d_{th})>0$ in this setup. 
In practice, it is not clear whether $H(d_{emp},d_{th})$ is non-zero solely because of sampling error or because the probability law of $d_{emp}$ is not equal to $d_{th}$. 
We correct the initial distance $H(d_{emp},d_{th})$ to virtually remove the finite sampling-based error and possibly allow the distance to be zero. Let us define $d_{RV}$ to be the histogram obtained from randomly generating $M$ pairs of random variables based on the theoretical PDF $d_{th}$. 
We then define the corrected Hellinger distance
\begin{eqnarray}\label{Hellinger_corrected}
 H_{corr}=H(d_{emp},d_{th})-\langle H(d_{RV},d_{th})\rangle_{RV},
\end{eqnarray}
where $\langle H(d_{RV},d_{th})\rangle_{RV}$ is the average Hellinger distance correction.  
This corrected distance can be very close to zero even at finite $M$. The average Hellinger distance correction is obtained by generating $p$ sets of $M$ random pairs, each with probability law according to $d_{th}$. Then the Hellinger distance is computed from each set and averaged by computing
\begin{equation}
    \langle H(d_{RV},d_{th})\rangle_{RV}=\frac{1}{p}\sum_{i=1}^p H(d^{\,(i)}_{RV},d_{th}).
\end{equation}
\subsection{Gaussian approximation}
\begin{figure}[h!]
\includegraphics[width=0.5\textwidth]{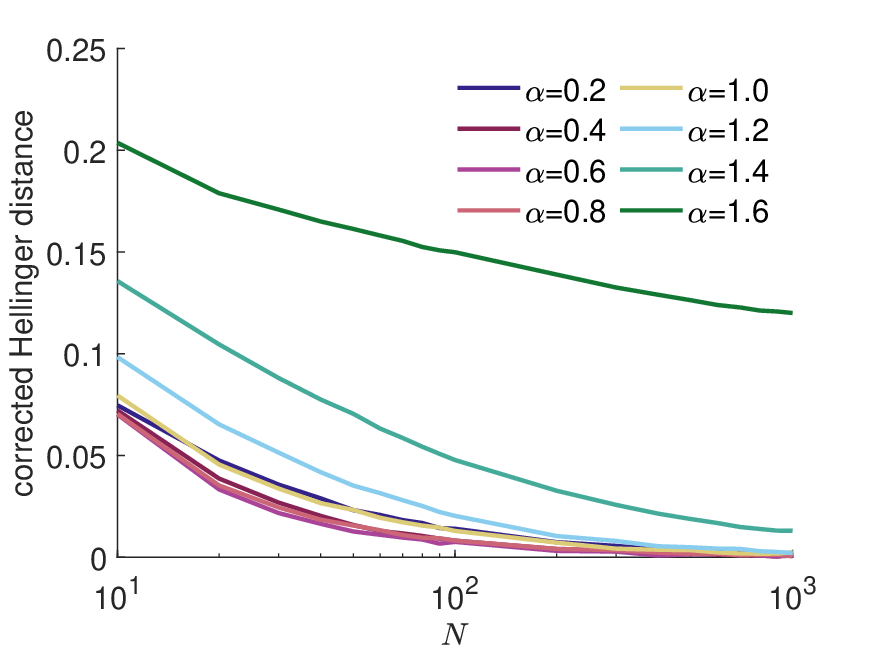}
\caption{Corrected Hellinger distance between theoretical joint PDF $p(\hat{\alpha},\hat{D})$ (using Gaussian approximation from Eq.~(17) in the main text) and the empirical joint PDF of $\hat{\alpha}$ and $\hat{D}$. Results are presented as a function of the trajectories' length $N$  received from $M=10^5$ fBm-particles for different values of the anomalous exponent $\alpha$ with
fixed $D=1$, $\Delta t=1$, and $\tau_{max}=5$.}
\label{fig:beta_fit_exponent_nonoise_constant_D1}
\end{figure}

To examine how the length of the trajectory $N$ influences the Gaussian approximation of the joint PDF, in Fig.~\ref{fig:beta_fit_exponent_nonoise_constant_D1} we show the corrected Hellinger distance as a function of $N$ between the theoretical joint PDF $p(\hat{\alpha},\hat{D})$ from Eq.~(17) in the main text and the empirical joint PDF of $\hat{\alpha}$ and $\hat{D}$ obtained from the fBm trajectories for different values of $\alpha$ with $D=1$.  The analysis is conducted for $M=10^5$ fBm trajectories assuming $\Delta t=1$. This corrected Hellinger distance can be considered as a quantitative measure to assess how well the Gaussian-based approximation explains the data.  As shown in Fig.~\ref{fig:beta_fit_exponent_nonoise_constant_D1}, the Gaussian approximation of the joint PDF $p(\hat{\alpha},\hat{D})$ is below $5$\% for $N=20$ when $\alpha\leq 1$. From $N=30$ onwards, the Gaussian approximation is considered very accurate, except for large values of $\alpha$, which aligns with the results presented in Fig.~1 in the main text.

\subsection{Copula approach beyond Gaussian approximation}
\begin{figure}[!htbp]
\includegraphics[width=0.4\textwidth]{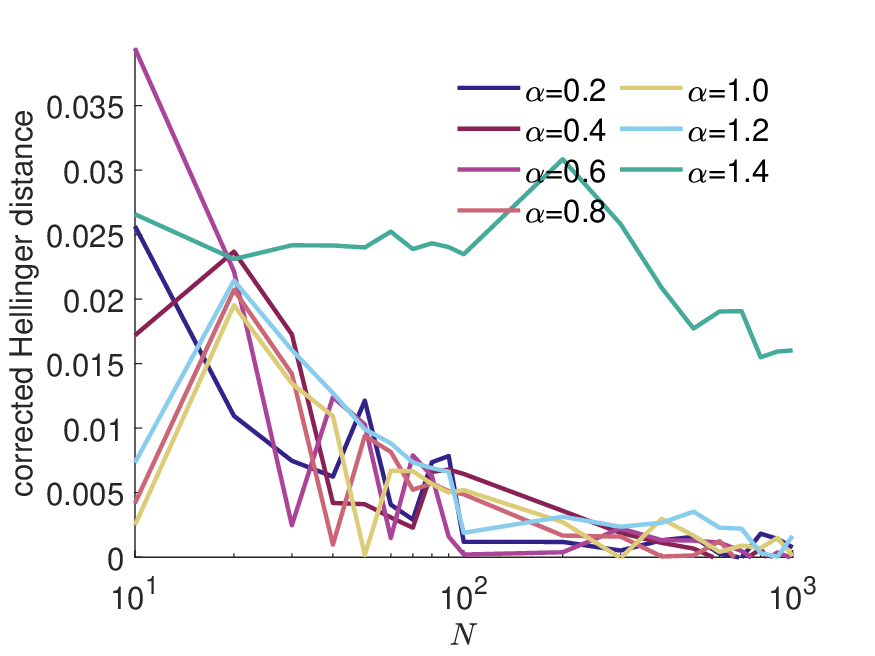}
 \caption{Corrected Hellinger distance between empirical joint pdf $p(\hat{\alpha},\hat{D})$ and theory for different values $\alpha$ obtained from a four moment approximation of the marginal PDFs of $\hat{\alpha}$ and $\hat{Y}$.  \label{fig:Hellinger_4_moments}}
 \end{figure}

In Fig.~\ref{fig:Hellinger_4_moments}, we show the results of the corrected Hellinger distance when the marginal PDFs of $\hat{\alpha}$ and $\hat{Y}$ are approximated on the basis of their empirical expectation, variance, skewness, and kurtosis.

The PDF of $\hat{\alpha}$ is a 4 4-parameter Beta distribution, and the $\hat{Y}$ is the Pearson type IV. The 4-moment approximation yields corrected Hellinger distances that are twice as small as those with the Gaussian approximation, but the cost is higher as skewness and kurtosis need to be computed. 

\section{Fitting of simulated data}
\begin{figure*}[!htbp]
\includegraphics[width=0.9\textwidth]{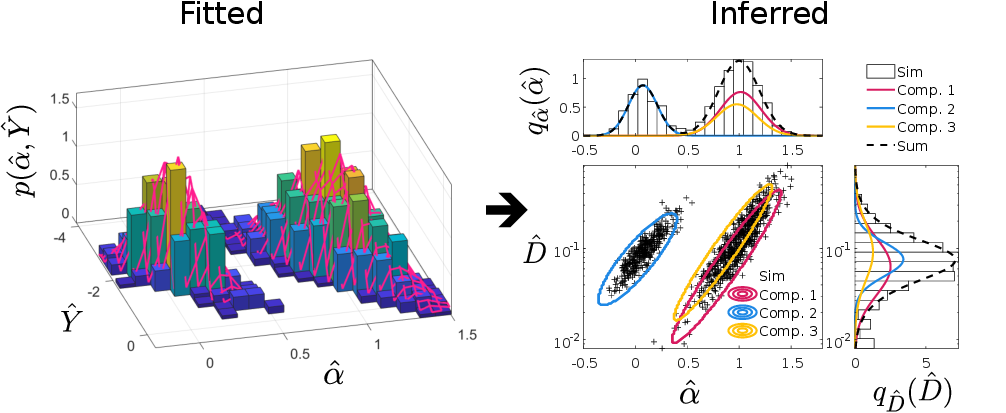}
 \caption{
Left panel: Histogram of $p(\hat{\alpha},\hat{Y})$ for a fBm with $M=1000$ trajectories of length $N=50$ and parameters $\alpha$ and $D$ drawn from density $p(\alpha, D)$ stated in Eq.~(19) from the main text with three components and compared versus the fitted distributed (pink mesh) by using Gaussian approximation. 
 Right panel: The histogram at the top of the panel displays the empirical marginal PDF $q_{\hat{\alpha}}(\hat{\alpha})$ with the corresponding marginal PDF of the three fitted components. The histogram at the right of the panel displays the empirical marginal PDF $q_{\hat{D}}(\hat{D})$ with the corresponding marginal PDF of the three fitted components. In the central plot, 
the data points $\hat{D} \, \textrm{vs} \, \hat{\alpha}$ are shown
with colored loops for each component delimiting when their joint PDF has a value of at least $0.5$.
All data are generated from the same ensemble of $M=10^3$ trajectories of fBm with parameters 
randomly picked among three components $\alpha_1=0.1$, $D_1=0.1$, $\alpha_2=1$, $D_2=0.1$, $\alpha_3=1$, $D_3=0.15$ with probability $c_1=c_2=c_3=1/3$, respectively.
The resulting fitted physical parameters by 
using the provided code are $\alpha_1=0.093$, $D_1=0.1$, $c_1=0.29$ (red lines); $\alpha_2=1$,
$D_2=0.099$, $c_2=0.31$ (blue lines); $\alpha_2=1$, $D_2=0.14$, $c_2=0.32$ (yellow lines).
}
\label{fig:beta_fit_exponent_demo}
\end{figure*}

Here we provide an example of the fitting of the estimated $p(\hat{\alpha},\hat{Y})$ to uncover the underlying joint PDF $p(\alpha,D)$ when it is a discrete three-component mixture with the same parameters as in Fig.~\ref{fig:joint_pdf_mixture_emp_vs_theory}. In contrast with Fig.~\ref{fig:joint_pdf_mixture_emp_vs_theory}, the analysis is performed over $M=10^3$ trajectories to have a reasonably low amount of data. The left image shows the histogram of the estimated $\hat{\alpha}$ and $\hat{Y}$ overlayed with a pinkish mesh, which is the fitted distribution. 
The right-hand side of
Fig.~\ref{fig:beta_fit_exponent_demo} compares the empirically obtained distributions with the fitted ones. 
The top histogram displays the empirical marginal PDF of 
$\hat{\alpha}$ overlayed with the individual marginal PDF of each component deduced from the fitting of the left image. The right histogram is that of $\hat{D}$ overlayed with the individual marginal PDFs of each component. Finally, the center image is the empirical plot of $\hat{D}$ as a function of $\hat{\alpha}$ containing one point per trajectory. Each contour delimits the range beyond which the joint PDF of each component has a probability density 
$\geq 0.5$.

\clearpage
\bibliographystyle{apsrev4-2} 


\end{document}